\documentclass[twocolumn,aps,showpacs]{revtex4}
\usepackage[dvips]{graphicx}
\usepackage{dcolumn}
\usepackage{bm}
\begin{document}

\title{What does measure the scaling exponent of the correlation sum in the case of human heart rate?}

\author{M.\ S\"akki, J.\ Kalda}

\affiliation{
Institute of Cybernetics,
Tallinn Technical University,
Akadeemia tee 21,
12618 Tallinn,
Estonia
}
\author{M.\ Vainu}
\affiliation{Tallinn Diagnostic Center, P\"arnu mnt.\ 104, Estonia}
\author{M.\ Laan}
\affiliation{N\~omme Children Hospital, Laste 1, Tallinn, Estonia}

\begin {abstract} 
It is shown that in the case of human heart rate,
the scaling behaviour of the correlation sum
(calculated by the Grassberger-Procaccia algorithm) 
is a result of the interplay of various factors:
finite resolution of the apparatus (finite-size effects),
a wide dynamic range of mean heart rate,
the amplitude of short-time variability being a decreasing function of the mean heart rate.
The value of the scaling exponent depends on all these factors and
is a certain measure of short-time variability of the signal.

\end {abstract} 

\maketitle

Heart rate variability (HRV) is often thought to be driven by 
deterministic chaos inside the heart. 
This chaos is explained in terms of dynamics of the
complex of sino-atrial and atrio-ventricular nodes, which has been 
successfully modeled as a system of non-linear coupled oscillators, responsible for
the heart rhythm \cite{Glass,West}.

As a consequence, correlation 
dimension and related quantities (like Lyapunov exponents and Kolmogorov entropy, etc.) 
have been thought to be important non-linear measures of HRV.
In particular, Babloyantz and Destexhe concluded 
\cite{Babloyantz} that high values of the correlation dimension indicate the 
healthiness of the heart. Later, the correlation dimension of the heart rate signal
has been calculated in a vast number of papers. 
Meanwhile, it has been pointed out that physiological time-series are 
typically non-stationary and noisy, and therefore, the correlation 
dimension cannot be calculated reliably \cite {Kantz,Kanters,Bezerianos} --- the fact which is
now widely accepted.
In the case of human heart,
the ``noise'' comes from the autonomous nervous system in the form of inputs regulating the 
heart rate (cf.\ \cite {Berne,Kaplan,Rosenblum}). These mostly non-deterministic signals suppress 
almost completely the underlying deterministic signal. 
Futhermore, it has been emphasized that a reasonable fitting of a correlation sum to a power law
does not necessarily mean that the obtained exponent is the correlation dimension
of the underlying dynamical system; instead, thorough non-automatable verification procedure has 
to be done \cite{KantzTextbook}.
All this leads us to the conclusion that
the formally calculated correlation dimension of a heart rhythm does not correspond to the dimensionality of 
an intrinsic attractor.

Thus, there are two important observations:
{\em (a)} the correlation sums of human heart rate follow typically a scaling law;
{\em (b)} in most cases, the scaling exponents are not the correlation dimensions.
Then, a natural question arises, what is the physical meaning
of these formally calculated exponents?

Our answer to the posed question is based on very simple observations, which are valid for
healthy patients: {\em(a)} the 
long-time variability of the inter-beat intervals (around 500 ms) is
typically much higher than the variability on the time scale of few heart beats (less than 50 ms);
{\em(b)} for those periods, when the mean heart rate is high (i.e.\ when the subject is
performing physical exercise) the heart rate variability is low;
{\em(c)} the heart rate is controlled  by non-deterministic and effectively random
signals arriving from the autonomous nervous system.
As a consequence, in time delay coordinates, an HRV time-series generates a 
baseball bat-shaped cloud of points.
Although the theoretical value of the correlation dimension of such a cloud is infinite, the
finite resolution of the recording apparatus, finite length of the time-series, and the linear 
structure of the cloud result in a  smaller value. This is evident for a very narrow ``bat'', 
which is efficiently one-dimensional. 
In what follows we show that the correlation dimension reflects the geometrical size of the 
cloud of points.

The layout of the paper is as follows. First, we give the details of the HRV database 
used for this study. Second, we provide a short overview of the research 
results related to the correlation dimension of human heart rhythm. Third, we construct a 
simple model-time-series, the correlation sum of which scales 
almost identically to that of real HRV data. Finally, we discuss the universality and 
implications of our model.

The experimental data analyzed in this paper have been recorded at Tallinn Diagnostic Center.
The recordings of ambulatory Holter-monitoring (24 hours, approximately 100 000 data points) 
were obtained during regular diagnostical examinations and covered over 200 patients 
with various clinically documented diagnoses (including also many healthy patients).
The main groups of patients are shown in Table 1.
The resolving power of recordings was 6 ms (sampling rate of 180 Hz).
The diagnostics and data verification has been made by qualified cardiologist;
the data preprocessing included also filtering of artifacts and arrhythmias.
\begin {table}[!bth]
\begin {tabular}{|l|c|c|c|c|c|c|c|}
\hline
         & Healthy & IHD & SND  & VES & PCI & RR & FSK\\
 \hline
No.\ of patients & 103    & 8   & 11   & 16  & 7   & 11 & 6 \\
\hline
Mean age & $45.5$ & $65.4$ & $50.0$ & $55.9$ & $47.3$ & $55.5$ & $11.7$\\
\hline
Std.\ dev.\ of age & $20.5$ &  $11.4$ & $19.3$ &  $14.3$ &  $11.6$ &  $14.4$ &  $4.6$ \\
\hline
\end{tabular}
\caption{
Test groups of patients. Abbreviations are as follows: IHD - Ischemic Heart Disease (Stenocardia); SND - Sinus Node Disease; 
VES - Ventricular Extrasystole; PCI - Post Cardiac Infarction; RR - Blood Pressure Disease; 
FSK - Functional Disease of Sinus Node.}
\end{table}

The concept of correlation dimension, introduced by Grassberger and Procaccia 
\cite{Grassberger}, is designed to reflect the number of degrees of
freedom of a deterministic system (or the dimensionality of an attractor, which, 
in principle, can be fractal). For empirical time-series, the phase variables 
are typically not known. It is expected that the 
attractors in the phase space are topologically equivalent to 
the attractors in a reconstructed phase space with 
time-lag coordinates $\{ x(t), x(t+\tau), \ldots x[t+(m-1)\tau]\}$, 
as long as the embedding dimensionality $m$ (the dimensionality of the reconstructed 
phase space) exceeds the dimensionality of the attractor $D$; here
$x(t)$ is the signal, and $\tau$ is a reasonably chosen time lag.
This circumstance is exploited by the Grassberger-Procaccia method \cite{Grassberger} 
for the calculation of the correlation dimension.
To begin with, the second order correlation sum is defined as
\begin {equation}
C_2(r) =  \frac2{N(N-1)}\sum_{i<j} \theta(r - \mid \bm r_{i} - \bm r_{j} \mid),
\end {equation}
where $\theta(r)$ is the Heaviside function, and 
$\bm r_{i} = \{x(t_i), x(t_i+\tau), \ldots, x[t+(m-1)\tau]\}$, is a point in the 
reconstructed phase space, and $i=1, 2, \ldots , N$ counts the moments of discretized time.
For small $r$, the correlation sum is expected to scale as $C_2(r) \propto r^{D_2}$,
assuming that $D_2 < m$. The exponent $D \equiv D_2$ is called 
the {\em correlation dimension} of the system.

A non-linear dynamical system may be  chaotic and then the phase trajectory fills the entire phase space.
In that case, the correlation dimension $D_2$ is equal to the number of degrees of freedom (the dimensionality 
of the phase space minus the number of conservation laws). This is why 
$D_2$ is often considered as a measure of the complexity of the system.
Babloyantz and Destexhe \cite{Babloyantz} studied the correlation dimension of the sequence of 
NN-intervals (intervals between normal heartbeats) of human heart rhythm. For healthy patients and data series consisting 
of 1000 intervals, they found $D=5.9\pm 0.4$. 
It is widely recognized that life threatening heart pathologies lead to 
the reduction of the complexity of the HRV signal, c.f.\ \cite{Bassingthwaighte}.
Correspondingly, the correlation dimension of the heart rate has been often considered
as a measure for the healthiness of the heart.

However, the heart is not an isolated system. Although the heart rhythm 
is generated by the complex of oscillatory elements, its rate is controlled by 
{\em non-deterministic inputs} arriving from the autonomous nervous system. 
In particular, these inputs lead to the increase of heart rate when the subject is 
under physical stress, and to slowing down when the subject is at rest, c.f.\ \cite{Kaplan}.
Healthy heart responds easily to these signals, and is able to adapt to a wide range 
of beating rates. This responsiveness gives rise to the high variability of the heart rate.
Severe heart diseases decrease the responsiveness 
of the heart with respect to the whole spectrum of signals arriving from the 
autonomous nervous system; this leads to the loss of
apparent complexity of the HRV signal.

The heart is more responsive  with respect to the signals of the autonomous nervous
system when the heart rate is slow, i.e.\  when the patient is at rest.
In that case, the heart rate variability is driven by weaker signals, like 
the ones generated by respiration and blood-pressure oscillations. These two stimuli are
quasi-periodic, the periods being  respectively a few and 10--20 seconds.
It should be noted that respiration can be mode-locked to the heart rate.
This mode-locking has been demonstrated by simultaneous recording of ECG and 
respiration activity, together with the technique called  cardiorespiratory synchrogram \cite{Kurths}. 
The ratio of the mode-locked periods can be small, 2:1, 3:1, 5:2, etc.,
and the phenomenon can give rise to certain patterns in the reconstructed 
phase space. These patterns can be easily misinterpreted as 
traces of an attractor of a non-linear deterministic system, therefore we discuss
this aspect in more details.

As mentioned above, HRV signals generate a baseball-bat-shaped
cloud of points in a time-lag space. For certain patients, the presence of less densely 
populated satellite clouds can be observed, see Fig.\ 1 (note that due to the sparse 
population of these clouds, the presence or absence of the clouds has almost no effect on the value of $D$).
We were able to show that for all such patients of our database,
the satellite clouds were not related to a deterministic chaos inside the heart. To this end, 
we studied the fluctuation function 
\begin{equation}
F(\nu ) = \left< |t_n - t_{n+\nu}|\right>
\end{equation}
(angular braces denote averaging over $n$).
Unlike in the case of ``single-cloud-patients'', the fluctuation function of the patients with satellite clouds 
revealed a presence of an oscillatory component, see Fig.\ 2a. By dividing the
entire 24-hour HRV record into one-hour intervals, and calculating the amplitude of the
oscillatory component of the fluctuation function for each interval, we were able to
locate the periods responsible for the satellite clouds in the reconstructed
phase space, see Fig.\ 2b. These were always the periods before falling asleep, 
around 10 or 11 pm, characterized by a low heart rate and a high respiration-driven
short-time variability. The phase between the heart rate and respiration is 
locked during tens of seconds, confirming the observations of Kurths et al.\ \cite{Kurths}.
Thus, in a certain sense, the heart and respiratory complex act as a system of coupled 
oscillators  c.f.\ \cite{Janson}; however, there is no determistic chaos inside the heart.
Note that our method of mode-locking detection is very simple, does not require synchronous
respiration rhythm recording (unlike the thourough method \cite{Kurths}), and can be conviniently used to find relatively short ($\agt$ 10\,min) locking periods 
from a 24-hour recording.
\begin {figure}[!bth]
\includegraphics{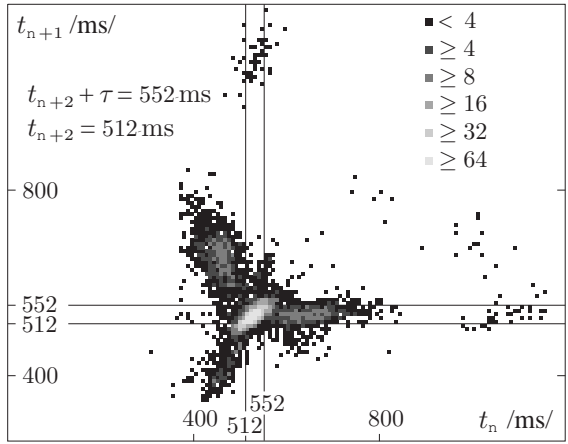}
\caption{Two-dimensional intersection of 3-dimensional reconstructed phase space 
for a patient with pronounced mode-locking between heart rate and respiration. 
The number of points per unit cell is given in gray-scale coding.}
\end{figure}
\begin {figure}[!bth]
\includegraphics*{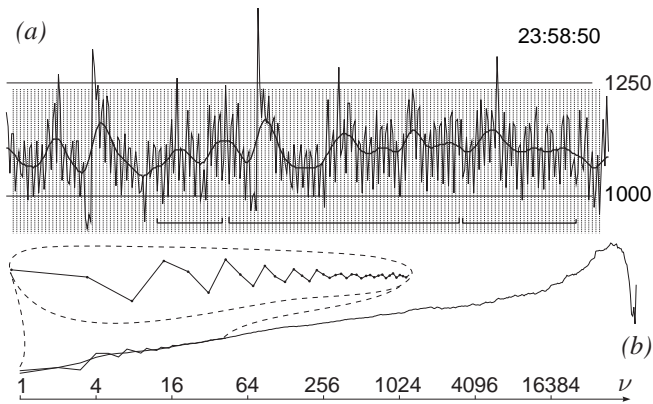}
\caption{Patient with 3:1 mode locking between heart rate and respiration:
{\em (a)} heart beat intervals (in milliseconds) plotted versus the beat number.
Heart rate has a pronounced oscillatory component;
vertical lines mark the period of three heart beats,  horizontal lines indicate the 
sequences  with coherent phase. {\em (b)}
Fluctuation function (arbitrary units) is plotted versus the time lag $\nu$ (in heart beats); the 
oscillating component is magnified.}
\end{figure}

Our model of the heart rhythm generation is as follows.
The non-linear deterministic dynamics inside the
heart is almost completely suppressed by the signals arriving from the 
autonomous nervous system. These signals control the mean heart rate, but 
obey also a noise-like component, the amplitude of which decreases with 
increasing mean heart rate. 
In the case of correlation dimension analysis, this
noise-like component is indistinguishable from a Gaussian noise.
Therefore, theoretically, the correlation dimension is infinite. The
reported relatively small values of the correlation dimension are to be 
attributed to the finite length of the time series and, most importantly, by finite resolution of
the recording apparatus. Regarding the length of the record: typically, the correlation dimension has been 
found  to be at the limit (or beyond) of a credible analysis \cite{Kantz,Smith}. Indeed, it has been 
suggested \cite{Kantz,Smith} that the calculation of the correlation dimension $D$ is reliable, 
if the number $N$ of data-points in the time series 
\begin{equation}
N \agt 10^{D/2+1}. 
\end{equation}
In Table 2, this criterion is compared with the data of some papers.
\begin {table}[!bth]
\begin {tabular}{|l|c|c|c|c|}
\hline
         & Ref. \cite{Babloyantz} & Ref. \cite{Kanters} & Ref. \cite{Govindan} & Ref. \cite {Guzzetti}\\
 \hline
Correlation dimension &   5.5--6.3  & 9.6--10.2   &  2.8-5.8 & 4--7\\
\hline
Length of the data set& $10^3$ & $2\cdot 10^4$ & $10^4$ &  $2\cdot 10^4$\\
\hline
Required length & $10^4$ &  $10^6$ & $10^4$ & $3\cdot 10^4$ \\
\hline
\end{tabular}
\caption{Data from papers devoted to the correlation dimension analysis:
experimental values of correlation dimension, lengths of the underlying data sets,
and data-set lengths required by Eq.\ 3.}
\end{table}

In order to test our hypothesis we aimed to construct such random time series
(using an algorithm as simple as possible), the correlation sum of which 
is similar to the correlation sums of the time series of real patients.
First we analyzed the sequences of NN-intervals extracted from ECG recordings.
The scaling exponent was calculated according 
to the Grassberger-Procaccia algorithm.
The six-dimensional embedding phase space was used for calculations.
The choice of the embedding dimensionality was motivated as follows.
To begin with, the analysis for phase space with $m > 6$ is not 
reliable due to sparse clouds of points (see Eq.\ 3). Further,
$m = 6$ does still make sense, because most of the previous studies have 
reported $D \alt 6$. 

Reliable correlation sum analysis is possible only for 
more or less stationary time series, cf.\ \cite{Kantz}.
Meanwhile, HRV signal is highly non-stationary, as is evidenced by the multifractal
structure of its long-time dynamics \cite{Ivanov}.
The most stationary period in the heart rate dynamics is the sleeping time.
This is why we studied only the nocturnal part of the HRV records.
The scaling exponent was determined as the slope of the correlation sum $C_2(r)$
in log-log plot by performing root-mean-square fit for the almost linear part 
(at small values of $r$) of the curve, see Fig.\ 3. Note that the leftmost 
horizontal  part of the curve is due to the 
limited resolving power (6 ms) of the medical equipment:
if two NN-intervals differ less than 6 ms, they are recorded to be of the same length.
The scaling exponents ranged from $D= 4.2$ to $D=5.1$ and were almost
uncorrelated with the diagnoses (see Table 2). 
\begin{table}[bth]
\includegraphics{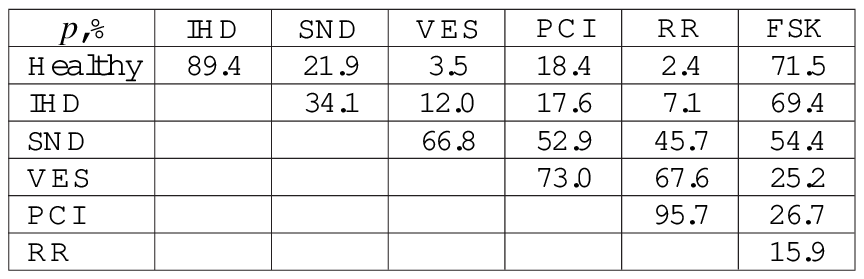}
\caption{$p$-values of the Student test for the seven group of patients. 
Abbreviations are explained in Table 1.}
\end{table}

\begin {figure}[!b]
\includegraphics{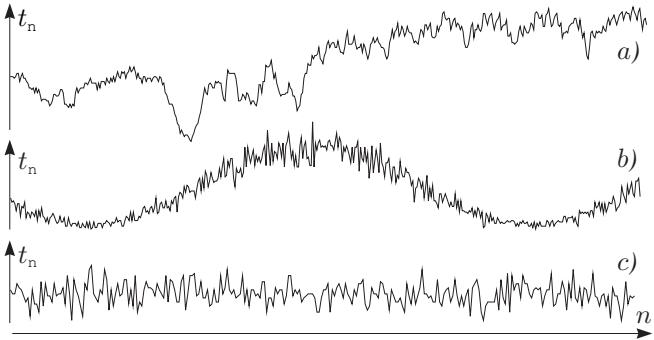}
\caption{Heart beat intervals (in arbitrary units) are plotted versus the beat number: 
{\em (a)} a real patient; {\em (b)} surrogate data (modulated
Gaussian noise); {\em (c)} plain Gaussian noise added to a constant ``heart'' rate.}
\end{figure}
Further we generated two surrogate time-series with Gaussian noise: {\em (i)} 
plain Gaussian noise added to a mean rate (see Fig.\ 4c); {\em (ii)}  time-series 
with variable mean rate and modulated noise, generated according to the formula
\begin {equation}
t_n = a + b \sin(f n) + c g(n) \sqrt{1.1 + \sin(f n)},
\end {equation}
see Fig.\ 4b. Here, $t_n$ denotes the duration of $n$-th interval;
$g(n)$ is a random normally distributed value with zero mean and standard deviation equal to 1.
The term $b \sin(f n)$ models the variability of the  mean heart rate due to
physiological processes (physical activity, blood pressure oscillations, etc.).
The term $\sqrt{1.1 + \sin(f n)}$ reflects the empirical observation that 
the short-time variability of heart rhythm increases together with the mean heart
beat interval. Note that both the square-root and sinusoidal dependances are 
rather arbitrary, the model is not sensitive neither with respect to the
particular functional dependencies nor with respect to the modulation frequency $f$. 
The numerical values of these parameters  have been chosen as follows:
$a = 500\,$ms, $b = 110\,$ms, $f = 0.005$, $c = 3.5\,$ms; 
the values of $t_n$ were rounded to the nearest multiple of $6\,$ms (the ``resolving power'').

For a Gaussian signal, the correlation dimension is infinite and the scaling exponent
should be equal to the embedding dimension $m=6$. This is exactly what is observed
for plain unmodulated Gaussian time-series, see Fig.\ 5. However, for the 
noise of modulated amplitude, the finite size effects are significant. 
Depending on the parameters $b$ and $c$, 
the correlation sum in Fig.\ 6 can be almost indistinguishable from the 
ones of real patients, see Fig.\ 4. The scaling exponent $D$ of such time-series depends on 
the amplitude of the Gaussian noise and on the resolving power (which, unlike in the 
case of real apparatus, was also freely adjustable).
By adjusting the above defined parameters 
$b$, $c$, and the resolving power, we were able to obtain the values 
ranging from $D=4$ to $D=6$.
\begin {figure}[!t]
\includegraphics{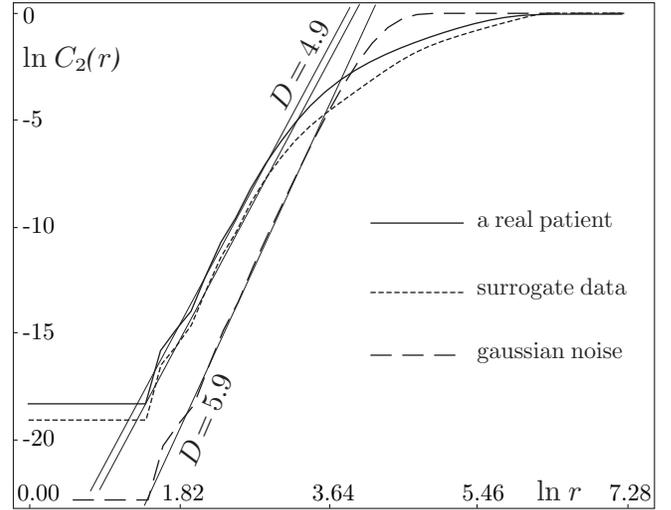}
\caption{Correlation sums of a healthy patient, a plain Gaussian signal, and a modulated Gaussian signal
in logarithmic plot. Embedding dimensionality $m=6$.}
\end{figure}

In conclusion, comparative analysis of real and surrogate data confirmed
that the scaling  behaviour of the correlation sum and
finite values of the formally calculated correlation dimension
$D$ are the result of the interplay of the following factors:
finite resolution of the recording equipment (which leads to finite-size effects),
a significant level of long-time variability (the dynamical range of the mean heart rate
exceeds the typical level of short-time variability),
the fact that the amplitude of short-time variability is a decreasing function of the mean heart rate.
The value of the scaling exponent is mostly defined by 
the dynamics of the short-time variability and resolving power of the recording apparatus; it
depends also on on the maximal embedding dimensionality.
Therefore, it can be used as a 
certain measure of short-time variability of the signal (however, in order to obtain comparable values, 
time-resolution and record length have to be kept constant). 
The diagnostic and/or prognostic value of this measure is possible, but 
found to be non-significant for our patient groups (see Table 2). 
We have also shown that the above drawn conclusion remains valid even in these cases,
when certain patterns (satellite clouds) can be
observed in time delay coordinates of heart rhythm (see Fig.\ 3).
These patterns are typically due to
the respiratory sinus arrhythmia and mode coupling between respiration and heart rhythm 
without any relation to the (possibly) derministic dynamics inside the heart.
Finally, we have devised a simple method of detecting the presence of this mode coupling,
based on fluctuation function (2).

The support of Estonian Science Foundation 
grant No.\ 4151 is acknowledged.
The authors are grateful to Prof.\ J.\ Engelbrecht for 
useful discussions.


\noindent
\end{document}